\documentclass[amsmath,amssymb,amsfonts,notitlepage,prl,aps,twocolumn,showpacs]{revtex4-1}

\usepackage{graphicx}

\begin{document}

\title{Strong coupling between single atoms and non-transversal photons}

\author{Christian Junge}
\author{Danny O'Shea}
\author{ J\"urgen Volz}
\author{Arno Rauschenbeutel}
\email{arno.rauschenbeutel@ati.ac.at}
\affiliation{Vienna Center for Quantum Science and Technology, Atominstitut, Vienna University of Technology, Vienna, Austria.}


\begin{abstract}
Light is often described as a fully transverse-polarized wave, i.e., with an electric field vector that is orthogonal to the direction of propagation. However, light confined in dielectric structures such as optical waveguides or whispering-gallery-mode microresonators can have a strong longitudinal polarization component. 
Here, using single $^{85}$Rb atoms strongly coupled to a whispering-gallery-mode microresonator, we experimentally and theoretically demonstrate that the presence of this longitudinal polarization fundamentally alters the interaction between light and matter.
\end{abstract}

\maketitle

The interaction between light and matter underlies basically every optical
process and application. For essentially plane waves in isotropic media, it
has been quantitatively investigated in a number of groundbreaking
experiments at the level of single atoms and single photons in high-finesse
cavities
\cite{Birnbaum:2005ab,Boozer:2007aa,Wilk:2007aa,Terraciano:2009aa,Kampschulte:2010aa,Volz:2011aa,Sayrin2011,Ritter:2012aa}. In order to further enhance the light--matter
coupling strength, an increasing number of recent experiments rely on waveguide structures \cite{Stiebeiner2009,Vetsch:2010aa,Hwang2011},  or high NA objectives \cite{Tey2008,Lee2011,Maiwald2012}. However, in these situations, the physics changes
drastically from the plane wave case because the polarization of the light
fields is in general no longer transversal but exhibits a longitudinal
component in the direction of propagation. This tags the propagation
direction of the light by its polarization state and fundamentally renders
full destructive interference of two counter-propagating waves impossible.
One would thus expect this effect to have
striking consequences for the physics of light--matter interaction. 

Here, we quantitatively investigate this phenomenon in a model system consisting of single
atoms that strongly interact with a whispering-gallery-mode (WGM)
microresonator \cite{Matsko:2006aa}. These resonators confine
light by continuous total internal reflection and offer the advantage of
very long photon lifetimes in conjunction with near-lossless in- and
out-coupling of light via tapered fibre couplers \cite{Spillane:2003aa}. As
recently demonstrated in a series of pioneering experiments with toroidal
WGM microresonators \cite{Aoki:2006aa,Park:2006aa,Srinivasan:2007aa,Dayan:2008aa,Aoki:2009aa,Alton:2010aa}, single atoms as well as solid state quantum emitters can be
strongly coupled to WGMs. 
Beyond their importance in strong light--matter coupling, WGM microresonators are highly versatile photonic devices which have found applications in a large variety of disciplines. They have enabled, e.g., on-chip detection of single nanoparticels \cite{Zhu:2010ab} and single viruses \cite{Vollmer:2008ab}, the generation of optical frequency combs \cite{DelHaye:2007aa} as well as squeezed and correlated twin-beams and single and pair photons \cite{Furst:2011aa,Fortsch:2012aa}. Moreover, WGM microresonators provide a successful experimental platform in the thriving field of cavity optomechanics \cite{Kippenberg:2008aa,Bahl:2012aa}.

\begin{figure}[h]
\includegraphics[width=7.5cm] {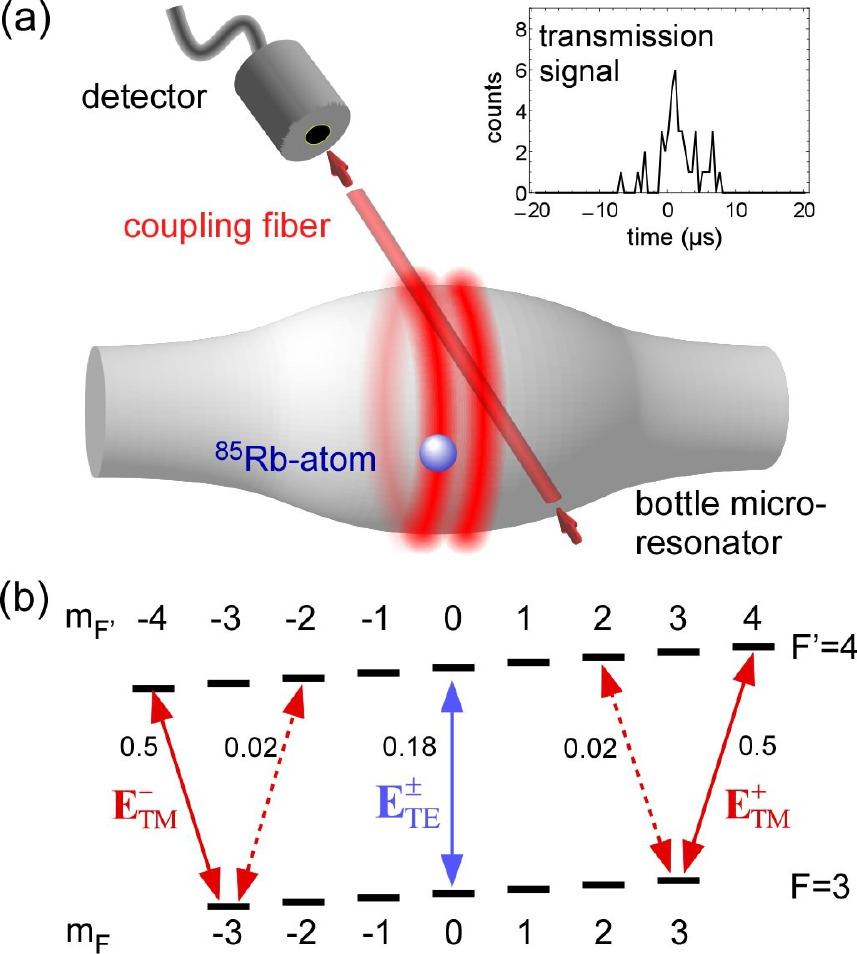}
\caption{
a) Schematic view of our bottle microresonator interfaced with a tapered fibre coupler. Probe light, launched into the coupling fibre, is  coupled into the resonator via frustrated total internal reflection and the remaining transmission is detected by a single photon counter. b) Relevant atomic levels and transition strengths for $^{85}$Rb. Due to the strong overlap with $\sigma^+$ ($\sigma^-$) polarization, TM modes drive $\Delta m_F=+1$ ($\Delta m_F=-1$) transitions, pumping into the Zeeman sublevel with maximum (minimum) $m_F$, while the $\pi$-polarized TE modes drive  $\Delta m_F=0$ transitions.}\label{fig1}
\end{figure}

However, thus far, the non-transversal polarization of WGMs has not been taken into account in the description of the quantum mechanical interaction of light and matter. In particular, WGM microresonators were conceptually treated as conventional ring resonators which sustain a pair of degenerate, identically polarized, counter-propagating modes \cite{Srinivasan:2007aa,He:2011aa,Zhu:2010ab,Aoki:2006aa,Dayan:2008aa,Aoki:2009aa,Alton:2010aa,Domokos:2000aa,Rosenblit:2004aa, Srinivasan:2007ab,Shen:2009ab,Stern:2011aa}. Under this assumption, the corresponding standing wave modes exhibit a full azimuthal intensity modulation with a phase that can be chosen such that one mode has a node at the position of a given emitter. As a consequence, this mode thus cannot interact with the atom, thereby leading to a fundamental limit of the performance of WGM microresonators for coupling light and matter. For instance, due to this uncoupled standing-wave mode, a single emitter should not be able to modify the on-resonant resonator transmission by more than 25\% even with an arbitrary large coupling strength (cf. Figure \ref{fig4}a). 

Here, we show that this picture is in general inadequate: In the case of non-transversally polarized WGMs, the two propagation directions of the photons are correlated with two nearly orthogonal polarization states. As a consequence, counter-propagating photons  are distinguishable by their polarization and cannot interfere destructively,  a situation not encountered in Fabry-P\'erot or conventional ring resonators. In particular, this effect prevents the formation of any uncoupled mode. In addition, the resonator field is almost perfectly circularly polarized in the plane of propagation despite the linear polarization of the pump light.

We investigate the consequences of this effect for light-matter coupling using WGM bottle microresonators \cite{Sumetsky2004,Louyer:2005aa,Pollinger:2009aa},
i.e., prolate-shaped cylindrically symmetric silica structures, see Fig.~\ref{fig1}a. These resonators sustain WGMs with ultra-high quality factor ($Q\gtrsim10^8$) and small mode volume, compatible with the requirements of CQED in the strong coupling regime \cite{Louyer:2005aa,Pollinger:2009aa}. Compared to other types of ultra-high $Q$ WGM microresonators, such as microspheres \cite{Braginsky:1989aa} and microtori \cite{Armani:2003aa}, bottle microresonators have the additional advantage of being fully tunable \cite{Pollinger:2009aa}. Moreover, their mode geometry straightforwardly enables near-lossless simultaneous coupling of two independent tapered fibre couplers \cite{Pollinger:2010aa}.

\begin{figure}[h]
\includegraphics[width=7.5cm] {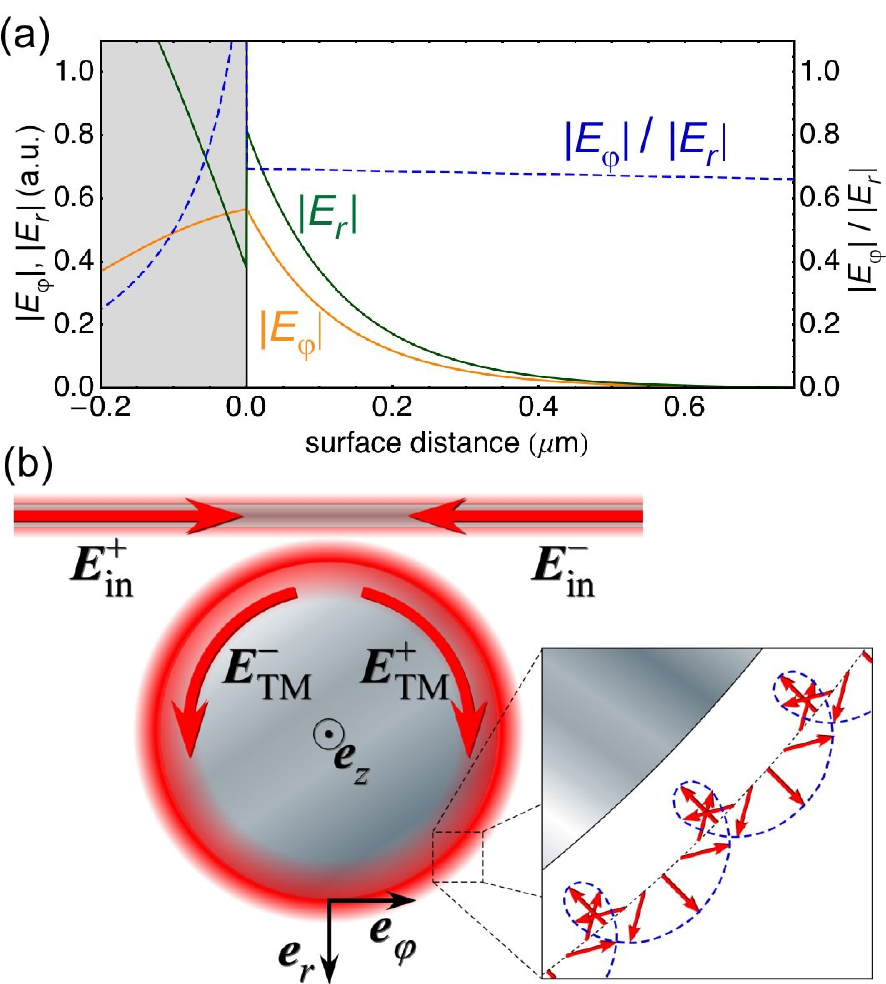}
\caption{
a) Longitudinal and transversal electric field components and their ratio as function of the distance to the resonator surface, calculated for the TM modes of our bottle microresonator. b) In the case of TM-modes, the spatial dependence of the electric field vector along the resonator is well described by a cycloid. }\label{fig2}
\end{figure}

In order to model the interaction of single atoms with bottle microresonator modes (bottle modes), we employ the Jaynes-Cummings Hamiltonian, generalized to a multilevel atom description and the full vectorial treatment of the evanescent electric field of a pair of counter-propagating bottle modes. As in all types of WGM microresonators, the light fields are guided by total internal reflection and the bottle modes can be classified according to the orientation of the electric field polarization, which either lies predominantly in the plane perpendicular to the resonator axis (TM) or predominantly points along this $z$-axis (TE). TE modes are almost exclusively transversally polarized, i.e., their electric field vector is, to a good approximation, perpendicular to their wave vector at any position of the mode. In qualitative contrast, the TM modes are non-transversally polarized \cite{Axelrod:1984aa,Kawalec:2007aa}, meaning that the electric field vector of the evanescent field has a non-vanishing component along the wave vector, see Fig.~\ref{fig2}a. This longitudinal component oscillates $\pm 90^\circ$ out of phase with respect to the transversal component. The $+$ ($-$) sign follows from Fresnel equations for  the left-handed (right-handed) propagation sense of the mode with respect to the $z$-axis, which defines our quantization axis. Thus, for a given distance from the resonator surface, the complex-valued amplitude vector of TM modes is
\begin{equation}\label{eq:polTM}
\mathbf{E}_{\rm TM}^\pm=\mathbf{E}_{\rm trans} \pm i\mathbf{E}_{\rm long}~.
\end{equation}
Here, $\mathbf{E}_{\rm trans}\approx|\mathbf{E}_{\rm trans}|\mathbf{e}_r$ ($\mathbf{E}_{\rm long} = |\mathbf{E}_{\rm long}|\mathbf{e}_\varphi$) is the transversal (longitudinal) amplitude vector with the real-valued radial (azimuthal) unit vector $\mathbf{e}_r$ ($\mathbf{e}_\varphi$). The situation is much simpler for the TE modes, which have approximately the same amplitude vector for both rotation senses, $\mathbf{E}_{\rm TE}^+\approx \mathbf{E}_{\rm TE}^-\approx|\mathbf{E}_{\rm TE}^\pm|\mathbf{e}_z$, where $\mathbf{e}_z$ is the real-valued axial unit vector.

The sign in Eq.~(\ref{eq:polTM}) has a decisive consequence for WGM resonators: For the sake of simplicity, let us first assume that $|\mathbf{E}_{\rm trans}|=|\mathbf{E}_{\rm long}|$. In this case, $\mathbf{E}_{\rm TM}^+$ and $\mathbf{E}_{\rm TM}^-$ describe  two modes with mutually orthogonal circular polarizations. However, in contrast to what is usually encountered for freely propagating light fields, the plane of polarization is parallel to the local wave vector of the mode and the electric field vector describes a cycloid along the circumference of the resonator, see Fig.~\ref{fig2}b. Any superposition of such a pair of counter-propagating modes corresponds to a light field with azimuthally symmetric intensity and ellipticity. In other words it is not possible to form an intensity modulated standing wave in the resonator.
 
As a consequence, if an atom is coupled to any superposition of degenerate counter propagating TM modes, the atom--light coupling strength is independent of the azimuthal position of the atom or, equivalently, of the relative phase of the superposition. In particular, this rules out the existence of an uncoupled standing wave mode with TM polarization. This differs fundamentally from what is encountered for standard paraxial Fabry-P\'erot and ring-resonators, where this intrinsic connection between polarization and propagation direction does not occur.

For WGMs with maximal angular momentum, the ratio $|\mathbf{E}_{\rm long}|/|\mathbf{E}_{\rm trans}|$ is largely independent of the geometry and the diameter of the WGM resonator and almost exclusively determined by the refractive index of the resonator material (see supplemental material). For WGM microresonators made of silica, one obtains $|\mathbf{E}_{\rm long}|/|\mathbf{E}_{\rm trans}|\approx 0.7$ (see Fig.~\ref{fig2}a). Thus, $|\mathbf{E}_{\rm TM}^\pm\cdot \mathbf{e}_{\sigma^\pm}^*|^2/|\mathbf{E}_{\rm TM}^\pm|^2>0.96$, where $\mathbf{e}_{\sigma^\pm}=(\mathbf{e}_r \pm i \mathbf{e}_\varphi)/\sqrt{2}$, meaning that $\mathbf{E}_{\rm TM}^\pm$-modes almost fully overlap with an ideally circularly ($\sigma^\pm$) polarized mode. For higher refractive indices, the overlap will be even higher.

\begin{figure}[h]
\includegraphics[width=7.5cm]{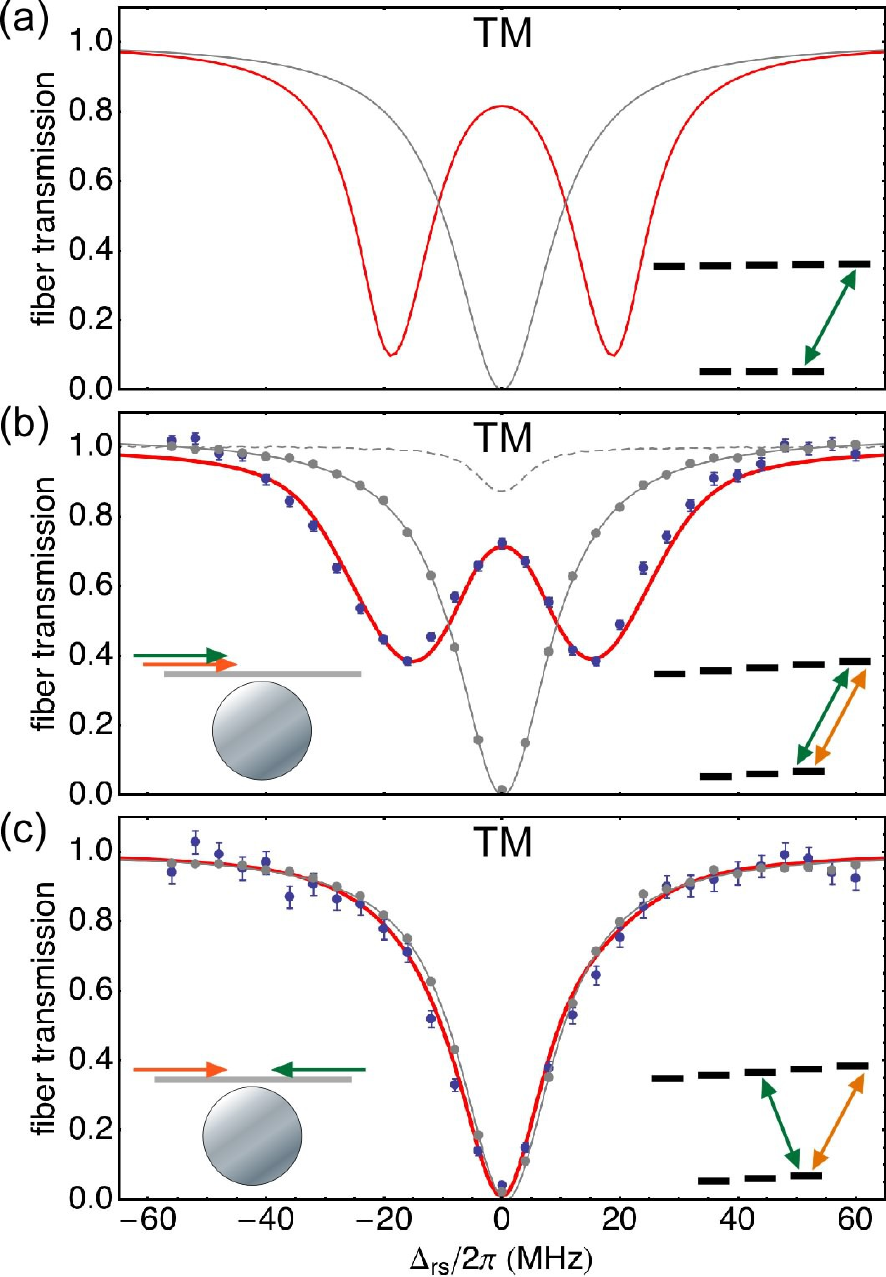} 
\caption{
a) Calculated coupling fibre transmission  for a TM mode coupled to a single atom with a fixed coupling strength of $g/2\pi=20$ MHz and zero atom--resonator detuning (red line). The transmission for  an empty resonator is shown by the grey line. b) Measured fibre transmission for our TM mode (blue circles: experimental data; error bars indicate the 1$\sigma$ statistical error) with a guiding field of $B_z=4.5$ G.  The red solid line is a theoretical fit yielding an average coupling strength of $\bar{g}/2\pi=17$ MHz with a standard deviation of $\sigma_g/2\pi=6$~MHz. The grey data shows the transmission of the empty cavity with a Lorentzian fit yielding a HWHM of $\kappa/2\pi=10$ MHz. The dashed line is the measured empty resonator transmission spectrum in the undercoupled regime. 
c) Vacuum Rabi spectrum measured under same conditions as in b) but applying the spectroscopy laser from the opposite direction as the detection laser. To avoid optical pumping by the spectroscopy laser, the spectrum is only measured for the first 100 ns. The theory curve is calculated using the same parameters $\bar{g}$ and $\sigma_g$ as in b). The insets show the direction of the detection (orange) and the spectroscopy light (green) and a simplified atomic level scheme indicating the driven transitions.}\label{fig3}
\end{figure}

\begin{figure}[h]
\includegraphics[width=7.5cm]{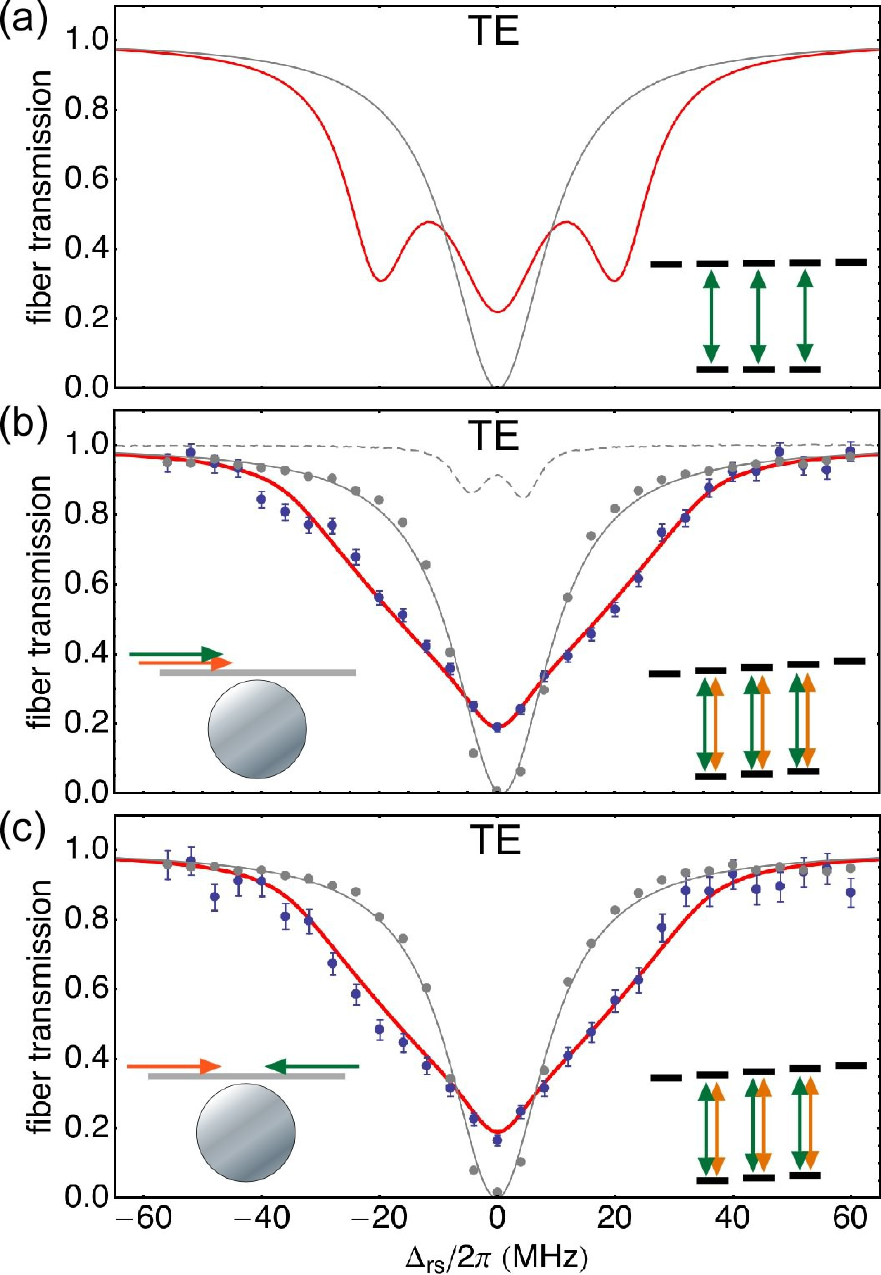} 
\caption{
a)-c) Same as Fig.~\ref{fig3} but for a TE polarized mode. The fit in b) yields an average coupling strength of $\bar{g}/2\pi=17$ MHz with a standard deviation of $\sigma_g/2\pi=9$~MHz. }\label{fig4}
\end{figure}

Besides preventing the formation of an uncoupled standing wave mode, the near perfect circular polarization of $\mathbf{E}_{\rm TM}^\pm$ has an additional advantage: Consider the case where the resonator is, e.g., coupled to a single $^{85}$Rb atom and light is launched into the coupling fibre in the direction of $\mathbf{E}^+_{\rm in}$ in order to resonantly excite mode $\mathbf{E}_{\rm TM}^+$, see Fig.~\ref{fig2}b. The atom then interacts with almost purely circularly polarized ($\sigma^+$) light and transitions with $\Delta m_F=+1$ are predominantly driven, where $m_F$ is the magnetic quantum number of the atomic hyperfine states. This leads to optical pumping and, when the $\mathbf{E}_{\rm TM}^+$-mode is close to resonance with the atomic transition $5S_{1/2},~F=3\to 5P_{3/2},~F^\prime=4$, the atom is transferred to the outermost Zeeman sublevel with $F=3,m_F=3$ after a few scattering events, see Fig.~\ref{fig1}b. From there, light in the $\mathbf{E}_{\rm TM}^+$-mode can only excite the $F^\prime=4, m_{F^\prime}=4$ state which can only decay back to the $F=3, m_F=3$ ground state. This situation occurs naturally and is highly advantageous for two reasons: First, the closed cycling transition maximizes the coupling strength of the atom to the electromagnetic field. 
Second, the selection rules for dipole transitions prevent the atom from emitting light into the orthogonally polarized, counter-propagating $\mathbf{E}_{\rm TM}^-$-mode. Despite the simultaneous existence of two degenerate resonator modes, this effectively leads to the ideal case of a two-level atom that only interacts with a single traveling-wave mode. Thus, our theoretical model predicts a spectrum of the coupled atom-resonator system that exhibits the well-known vacuum-Rabi splitting of the resonance frequencies by $\Delta\omega=2g$, see Fig.~\ref{fig3}a, where $g$ is the atom-resonator coupling strength.

In the case of TE-modes, both running waves have the same linear polarization. Thus, the atom will interact with both modes and an uncoupled standing wave occurs. This leads to qualitatively different physical behavior compared to the TM case,  which is apparent in the predicted spectrum which contains an additional central resonance, that is a signature of the uncoupled standing wave mode, see Fig.~\ref{fig4}a. This spectrum essentially corresponds to the predictions of the formerly employed theoretical model that neglected the non-transversal polarization of the resonator modes.

In order to verify that the above scenarios adequately describe our experiment, we investigate the physical behavior of TE and TM modes. Figure \ref{fig1}a shows a schematic of our experimental setup with the bottle microresonator interfaced with a sub-micron diameter coupling fibre \cite{OShea:2010aa,OShea:2011ab,Junge:2011aa}. We tune the frequency of the bottle microresonator to the atomic $5S_{1/2},~F=3\to 5P_{3/2},~F^\prime=4$ transition of $^{85}$Rb (transition wavelength $\lambda= 780$~nm) and critically couple the  coupling fibre to the resonator. At this set-point, the in- and out-coupling rate of light matches the intrinsic loss rate of the empty resonator. We set the polarization of the resonant light in the coupling fibre to match the polarization of the  bottle mode. As a result, the light will be entirely coupled into the bottle microresonator mode and dissipated therein. The remaining transmission through the coupling fibre is typically as low as 1~\%, see Figs.~\ref{fig3},\ref{fig4}. We then launch a cloud of laser-cooled $^{85}$Rb-atoms towards the resonator. If an atom enters the evanescent field of the resonator-mode, the vacuum Rabi splitting of the resonance frequency of the strongly coupled atom--resonator system results in a significant increase of the coupling fibre transmission, see inset of Fig.~\ref{fig1}a. When such an event is registered, a fast optical switch is used to turn off the detection light in realtime and to simultaneously apply a spectroscopy field through the coupling fibre that can be set to any relevant detuning $\Delta_{\rm rs}=\omega_{\rm r}-\omega_{\rm s}$, where $\omega_{\rm r}/2\pi$ and $\omega_{\rm s}/2\pi$ are the frequencies of the resonator mode and  spectroscopy field, respectively.

Figures \ref{fig3}b and \ref{fig4}b show transmission spectra obtained for the case of probing a TM- and TE-polarized mode, $\mathbf{E}_{\rm TM}^+$ and $\mathbf{E}_{\rm TE}^+$, respectively.  A guiding magnetic field $B_z=4.5$~G along the $z$-axis lifts the degeneracy of the Zeeman sublevels. The bottle-resonator is set to resonance with the dominant atomic transition ($F=3,~m_F=3\to F^\prime=4,~m_F^\prime=4$ for TM and $F=3,~m_F=0\to F^\prime=4,~m_F^\prime=0$ for TE). Compared to the case of an atom-resonator system with a well defined coupling constant $g$ (Fig. \ref{fig3}a and \ref{fig4}a), in our experiment the coupling strength of the atoms varies, both during the transit and from shot to shot. We account for this by fitting an averaged spectrum to the transmission data, where the only free fit parameters are the mean value  of the coupling strength $\bar{g}$ and its standard deviation $\sigma_g$. Perfect agreement is found between the experimental data and the theoretically predicted spectra. As expected, for the $\mathbf{E}_{\rm TM}^+$-mode, the spectrum clearly features two transmission dips while the spectrum for the $\mathbf{E}_{\rm TE}^+$-mode is  dominated by a central dip.

We now launch the detection light along $\mathbf{E}_{\rm in}^+$ and the spectroscopy light along the opposite direction $\mathbf{E}_{\rm in}^-$. For the TE polarized mode, we expect the same transmission spectrum as before. However, the situation is different for the TM modes, where we populate the $\mathbf{E}_{\rm TM}^+$-mode during detection and the atom is pumped into the $m_F=3$ state. The spectroscopy is then carried out on the $\mathbf{E}_{\rm TM}^-$-mode, which initially drives the $F=3,~m_F=3\to F^\prime=4,~m_F^\prime=2$ transition, see Fig.~\ref{fig1}b. This transition is strongly suppressed by two effects. First, its strength is more than one order of magnitude smaller than for the closed cycling transition, and second, the strong interaction of the atom with the empty $\mathbf{E}_{\rm TM}^+$-mode on the $F=3,~m_F=1\to F^\prime=4,~m_F^\prime=2$ transition leads to a vacuum Rabi splitting of the excited state. As a result, the transmission spectrum recorded with detection and spectroscopy light launched from opposite sides should initially closely resemble that of an empty resonator. 
Figure \ref{fig3}c (\ref{fig4}c) shows the transmission spectra of the TM (TE) mode recorded for this case. The interaction time of the spectroscopy field with the coupled atom--resonator system was limited to the first 100~ns after the initial relaxation of the resonator field.
We experimentally verified that this time interval is short enough to prevent optical depumping of the initial atomic state. The transmission spectrum predicted by our model in Figs. \ref{fig3}c  and \ref{fig4}c  (solid lines with the same values of $g$ and $\sigma$ as in Figs. \ref{fig3}b and \ref{fig4}b) shows excellent agreement between experiment and theory.

Summarizing, we experimentally and theoretically demonstrated that the longitudinal component of the electric field in whispering-gallery-modes intrinsically correlates the polarization of counter-propagating photons in the resonator with their propagation direction. 
Beyond whispering-gallery-mode resonators, this effect inevitably occurs in all physical systems and situations which involve longitudinal polarization components. In particular, this includes optical waveguides, nanophotonic structures and even focused beams that propagate in free space. The polarization--propagation correlation inhibits full destructive interference between counter-propagating photons and can thus qualitatively alter the resulting light field as to compared to what is expected when treating light as fully transversal wave.
In the case of WGM resonators, this phenomenon leads to a highly advantageous situation that allows one to overcome the limitations of traditional ring resonators for coupling light and matter. In particular, it enables the control of the interaction between the atom and the two counter-propagating modes in the resonator. In combination with the demonstrated low optical losses of whispering-gallery-mode microresonators, this constitutes ideal conditions for the realization of photonic quantum devices in optical fibre-based networks. 

We gratefully acknowledge financial support by the European Science Foundation and the Volkswagen Foundation. J.V.  acknowledges support by the European Commission (Marie Curie IEF Grant 300392). C.J. acknowledges support by the German National Academic Foundation.

\bibliography{literature}

\begin{thebibliography}{46}%
\makeatletter
\providecommand \@ifxundefined [1]{%
 \@ifx{#1\undefined}
}%
\providecommand \@ifnum [1]{%
 \ifnum #1\expandafter \@firstoftwo
 \else \expandafter \@secondoftwo
 \fi
}%
\providecommand \@ifx [1]{%
 \ifx #1\expandafter \@firstoftwo
 \else \expandafter \@secondoftwo
 \fi
}%
\providecommand \natexlab [1]{#1}%
\providecommand \enquote  [1]{``#1''}%
\providecommand \bibnamefont  [1]{#1}%
\providecommand \bibfnamefont [1]{#1}%
\providecommand \citenamefont [1]{#1}%
\providecommand \href@noop [0]{\@secondoftwo}%
\providecommand \href [0]{\begingroup \@sanitize@url \@href}%
\providecommand \@href[1]{\@@startlink{#1}\@@href}%
\providecommand \@@href[1]{\endgroup#1\@@endlink}%
\providecommand \@sanitize@url [0]{\catcode `\\12\catcode `\$12\catcode
  `\&12\catcode `\#12\catcode `\^12\catcode `\_12\catcode `\%12\relax}%
\providecommand \@@startlink[1]{}%
\providecommand \@@endlink[0]{}%
\providecommand \url  [0]{\begingroup\@sanitize@url \@url }%
\providecommand \@url [1]{\endgroup\@href {#1}{\urlprefix }}%
\providecommand \urlprefix  [0]{URL }%
\providecommand \Eprint [0]{\href }%
\providecommand \doibase [0]{http://dx.doi.org/}%
\providecommand \selectlanguage [0]{\@gobble}%
\providecommand \bibinfo  [0]{\@secondoftwo}%
\providecommand \bibfield  [0]{\@secondoftwo}%
\providecommand \translation [1]{[#1]}%
\providecommand \BibitemOpen [0]{}%
\providecommand \bibitemStop [0]{}%
\providecommand \bibitemNoStop [0]{.\EOS\space}%
\providecommand \EOS [0]{\spacefactor3000\relax}%
\providecommand \BibitemShut  [1]{\csname bibitem#1\endcsname}%
\let\auto@bib@innerbib\@empty
\bibitem [{\citenamefont {Birnbaum}\ \emph {et~al.}(2005)\citenamefont
  {Birnbaum}, \citenamefont {Boca}, \citenamefont {Miller}, \citenamefont
  {Boozer}, \citenamefont {Northup},\ and\ \citenamefont
  {Kimble}}]{Birnbaum:2005ab}%
  \BibitemOpen
  \bibfield  {author} {\bibinfo {author} {\bibfnamefont {K.~M.}\ \bibnamefont
  {Birnbaum}}, \bibinfo {author} {\bibfnamefont {A.}~\bibnamefont {Boca}},
  \bibinfo {author} {\bibfnamefont {R.}~\bibnamefont {Miller}}, \bibinfo
  {author} {\bibfnamefont {A.~D.}\ \bibnamefont {Boozer}}, \bibinfo {author}
  {\bibfnamefont {T.~E.}\ \bibnamefont {Northup}}, \ and\ \bibinfo {author}
  {\bibfnamefont {H.~J.}\ \bibnamefont {Kimble}},\ }\href@noop {} {\bibfield
  {journal} {\bibinfo  {journal} {Nature}\ }\textbf {\bibinfo {volume} {436}},\
  \bibinfo {pages} {87} (\bibinfo {year} {2005})}\BibitemShut {NoStop}%
\bibitem [{\citenamefont {Boozer}\ \emph {et~al.}(2007)\citenamefont {Boozer},
  \citenamefont {Boca}, \citenamefont {Miller}, \citenamefont {Northup},\ and\
  \citenamefont {Kimble}}]{Boozer:2007aa}%
  \BibitemOpen
  \bibfield  {author} {\bibinfo {author} {\bibfnamefont {A.~D.}\ \bibnamefont
  {Boozer}}, \bibinfo {author} {\bibfnamefont {A.}~\bibnamefont {Boca}},
  \bibinfo {author} {\bibfnamefont {R.}~\bibnamefont {Miller}}, \bibinfo
  {author} {\bibfnamefont {T.~E.}\ \bibnamefont {Northup}}, \ and\ \bibinfo
  {author} {\bibfnamefont {H.~J.}\ \bibnamefont {Kimble}},\ }\href {\doibase
  10.1103/PhysRevLett.98.193601} {\bibfield  {journal} {\bibinfo  {journal}
  {Phys. Rev. Lett.}\ }\textbf {\bibinfo {volume} {98}},\ \bibinfo {pages}
  {193601} (\bibinfo {year} {2007})}\BibitemShut {NoStop}%
\bibitem [{\citenamefont {Wilk}\ \emph {et~al.}(2007)\citenamefont {Wilk},
  \citenamefont {Webster}, \citenamefont {Kuhn},\ and\ \citenamefont
  {Rempe}}]{Wilk:2007aa}%
  \BibitemOpen
  \bibfield  {author} {\bibinfo {author} {\bibfnamefont {T.}~\bibnamefont
  {Wilk}}, \bibinfo {author} {\bibfnamefont {S.}~\bibnamefont {Webster}},
  \bibinfo {author} {\bibfnamefont {A.}~\bibnamefont {Kuhn}}, \ and\ \bibinfo
  {author} {\bibfnamefont {G.}~\bibnamefont {Rempe}},\ }\href@noop {}
  {\bibfield  {journal} {\bibinfo  {journal} {Science}\ }\textbf {\bibinfo
  {volume} {317}},\ \bibinfo {pages} {488} (\bibinfo {year}
  {2007})}\BibitemShut {NoStop}%
\bibitem [{\citenamefont {Terraciano}\ \emph {et~al.}(2009)\citenamefont
  {Terraciano}, \citenamefont {Olson~Knell}, \citenamefont {Norris},
  \citenamefont {Jing}, \citenamefont {Fernandez},\ and\ \citenamefont
  {Orozco}}]{Terraciano:2009aa}%
  \BibitemOpen
  \bibfield  {author} {\bibinfo {author} {\bibfnamefont {M.~L.}\ \bibnamefont
  {Terraciano}}, \bibinfo {author} {\bibfnamefont {R.}~\bibnamefont
  {Olson~Knell}}, \bibinfo {author} {\bibfnamefont {D.~G.}\ \bibnamefont
  {Norris}}, \bibinfo {author} {\bibfnamefont {J.}~\bibnamefont {Jing}},
  \bibinfo {author} {\bibfnamefont {A.}~\bibnamefont {Fernandez}}, \ and\
  \bibinfo {author} {\bibfnamefont {L.~A.}\ \bibnamefont {Orozco}},\
  }\href@noop {} {\bibfield  {journal} {\bibinfo  {journal} {Nat. Phys.}\
  }\textbf {\bibinfo {volume} {5}},\ \bibinfo {pages} {480} (\bibinfo {year}
  {2009})}\BibitemShut {NoStop}%
\bibitem [{\citenamefont {Kampschulte}\ \emph {et~al.}(2010)\citenamefont
  {Kampschulte}, \citenamefont {Alt}, \citenamefont {Brakhane}, \citenamefont
  {Eckstein}, \citenamefont {Reimann}, \citenamefont {Widera},\ and\
  \citenamefont {Meschede}}]{Kampschulte:2010aa}%
  \BibitemOpen
  \bibfield  {author} {\bibinfo {author} {\bibfnamefont {T.}~\bibnamefont
  {Kampschulte}}, \bibinfo {author} {\bibfnamefont {W.}~\bibnamefont {Alt}},
  \bibinfo {author} {\bibfnamefont {S.}~\bibnamefont {Brakhane}}, \bibinfo
  {author} {\bibfnamefont {M.}~\bibnamefont {Eckstein}}, \bibinfo {author}
  {\bibfnamefont {R.}~\bibnamefont {Reimann}}, \bibinfo {author} {\bibfnamefont
  {A.}~\bibnamefont {Widera}}, \ and\ \bibinfo {author} {\bibfnamefont
  {D.}~\bibnamefont {Meschede}},\ }\href@noop {} {\bibfield  {journal}
  {\bibinfo  {journal} {Phys. Rev. Lett.}\ }\textbf {\bibinfo {volume} {105}},\
  \bibinfo {pages} {153603} (\bibinfo {year} {2010})}\BibitemShut {NoStop}%
\bibitem [{\citenamefont {Volz}\ \emph {et~al.}(2011)\citenamefont {Volz},
  \citenamefont {Gehr}, \citenamefont {Dubois}, \citenamefont {Esteve},\ and\
  \citenamefont {Reichel}}]{Volz:2011aa}%
  \BibitemOpen
  \bibfield  {author} {\bibinfo {author} {\bibfnamefont {J.}~\bibnamefont
  {Volz}}, \bibinfo {author} {\bibfnamefont {R.}~\bibnamefont {Gehr}}, \bibinfo
  {author} {\bibfnamefont {G.}~\bibnamefont {Dubois}}, \bibinfo {author}
  {\bibfnamefont {J.}~\bibnamefont {Esteve}}, \ and\ \bibinfo {author}
  {\bibfnamefont {J.}~\bibnamefont {Reichel}},\ }\href@noop {} {\bibfield
  {journal} {\bibinfo  {journal} {Nature}\ }\textbf {\bibinfo {volume} {475}},\
  \bibinfo {pages} {210} (\bibinfo {year} {2011})}\BibitemShut {NoStop}%
\bibitem [{\citenamefont {Sayrin}\ \emph {et~al.}(2011)\citenamefont {Sayrin},
  \citenamefont {Dotsenko}, \citenamefont {Zhou}, \citenamefont {Peaudecerf},
  \citenamefont {Rybarczyk}, \citenamefont {Gleyzes}, \citenamefont {Rouchon},
  \citenamefont {Mirrahimi}, \citenamefont {Amini}, \citenamefont {Brune},
  \citenamefont {Raimond},\ and\ \citenamefont {Haroche}}]{Sayrin2011}%
  \BibitemOpen
  \bibfield  {author} {\bibinfo {author} {\bibfnamefont {C.}~\bibnamefont
  {Sayrin}}, \bibinfo {author} {\bibfnamefont {I.}~\bibnamefont {Dotsenko}},
  \bibinfo {author} {\bibfnamefont {X.}~\bibnamefont {Zhou}}, \bibinfo {author}
  {\bibfnamefont {B.}~\bibnamefont {Peaudecerf}}, \bibinfo {author}
  {\bibfnamefont {T.}~\bibnamefont {Rybarczyk}}, \bibinfo {author}
  {\bibfnamefont {S.}~\bibnamefont {Gleyzes}}, \bibinfo {author} {\bibfnamefont
  {P.}~\bibnamefont {Rouchon}}, \bibinfo {author} {\bibfnamefont
  {M.}~\bibnamefont {Mirrahimi}}, \bibinfo {author} {\bibfnamefont
  {H.}~\bibnamefont {Amini}}, \bibinfo {author} {\bibfnamefont
  {M.}~\bibnamefont {Brune}}, \bibinfo {author} {\bibfnamefont {J.-M.}\
  \bibnamefont {Raimond}}, \ and\ \bibinfo {author} {\bibfnamefont
  {S.}~\bibnamefont {Haroche}},\ }\href@noop {} {\bibfield  {journal} {\bibinfo
   {journal} {Nature}\ }\textbf {\bibinfo {volume} {477}},\ \bibinfo {pages}
  {73} (\bibinfo {year} {2011})}\BibitemShut {NoStop}%
\bibitem [{\citenamefont {Ritter}\ \emph {et~al.}(2012)\citenamefont {Ritter},
  \citenamefont {Nolleke}, \citenamefont {Hahn}, \citenamefont {Reiserer},
  \citenamefont {Neuzner}, \citenamefont {Uphoff}, \citenamefont {Mucke},
  \citenamefont {Figueroa}, \citenamefont {Bochmann},\ and\ \citenamefont
  {Rempe}}]{Ritter:2012aa}%
  \BibitemOpen
  \bibfield  {author} {\bibinfo {author} {\bibfnamefont {S.}~\bibnamefont
  {Ritter}}, \bibinfo {author} {\bibfnamefont {C.}~\bibnamefont {Nolleke}},
  \bibinfo {author} {\bibfnamefont {C.}~\bibnamefont {Hahn}}, \bibinfo {author}
  {\bibfnamefont {A.}~\bibnamefont {Reiserer}}, \bibinfo {author}
  {\bibfnamefont {A.}~\bibnamefont {Neuzner}}, \bibinfo {author} {\bibfnamefont
  {M.}~\bibnamefont {Uphoff}}, \bibinfo {author} {\bibfnamefont
  {M.}~\bibnamefont {Mucke}}, \bibinfo {author} {\bibfnamefont
  {E.}~\bibnamefont {Figueroa}}, \bibinfo {author} {\bibfnamefont
  {J.}~\bibnamefont {Bochmann}}, \ and\ \bibinfo {author} {\bibfnamefont
  {G.}~\bibnamefont {Rempe}},\ }\href@noop {} {\bibfield  {journal} {\bibinfo
  {journal} {Nature}\ }\textbf {\bibinfo {volume} {484}},\ \bibinfo {pages}
  {195} (\bibinfo {year} {2012})}\BibitemShut {NoStop}%
\bibitem [{\citenamefont {Stiebeiner}\ \emph {et~al.}(2009)\citenamefont
  {Stiebeiner}, \citenamefont {Rehband}, \citenamefont {Garcia-Fernandez},\
  and\ \citenamefont {Rauschenbeutel}}]{Stiebeiner2009}%
  \BibitemOpen
  \bibfield  {author} {\bibinfo {author} {\bibfnamefont {A.}~\bibnamefont
  {Stiebeiner}}, \bibinfo {author} {\bibfnamefont {O.}~\bibnamefont {Rehband}},
  \bibinfo {author} {\bibfnamefont {R.}~\bibnamefont {Garcia-Fernandez}}, \
  and\ \bibinfo {author} {\bibfnamefont {A.}~\bibnamefont {Rauschenbeutel}},\
  }\href {http://www.opticsexpress.org/abstract.cfm?URI=oe-17-24-21704}
  {\bibfield  {journal} {\bibinfo  {journal} {Opt. Express}\ }\textbf {\bibinfo
  {volume} {17}},\ \bibinfo {pages} {21704} (\bibinfo {year}
  {2009})}\BibitemShut {NoStop}%
\bibitem [{\citenamefont {Vetsch}\ \emph {et~al.}(2010)\citenamefont {Vetsch},
  \citenamefont {Reitz}, \citenamefont {Sagu{\'e}}, \citenamefont {Schmidt},
  \citenamefont {Dawkins},\ and\ \citenamefont
  {Rauschenbeutel}}]{Vetsch:2010aa}%
  \BibitemOpen
  \bibfield  {author} {\bibinfo {author} {\bibfnamefont {E.}~\bibnamefont
  {Vetsch}}, \bibinfo {author} {\bibfnamefont {D.}~\bibnamefont {Reitz}},
  \bibinfo {author} {\bibfnamefont {G.}~\bibnamefont {Sagu{\'e}}}, \bibinfo
  {author} {\bibfnamefont {R.}~\bibnamefont {Schmidt}}, \bibinfo {author}
  {\bibfnamefont {S.~T.}\ \bibnamefont {Dawkins}}, \ and\ \bibinfo {author}
  {\bibfnamefont {A.}~\bibnamefont {Rauschenbeutel}},\ }\href@noop {}
  {\bibfield  {journal} {\bibinfo  {journal} {Phys. Rev. Lett.}\ }\textbf
  {\bibinfo {volume} {104}},\ \bibinfo {pages} {203603} (\bibinfo {year}
  {2010})}\BibitemShut {NoStop}%
\bibitem [{\citenamefont {Hwang}\ and\ \citenamefont
  {Hinds}(2011)}]{Hwang2011}%
  \BibitemOpen
  \bibfield  {author} {\bibinfo {author} {\bibfnamefont {J.}~\bibnamefont
  {Hwang}}\ and\ \bibinfo {author} {\bibfnamefont {E.~A.}\ \bibnamefont
  {Hinds}},\ }\href {http://stacks.iop.org/1367-2630/13/i=8/a=085009}
  {\bibfield  {journal} {\bibinfo  {journal} {New Journal of Physics}\ }\textbf
  {\bibinfo {volume} {13}},\ \bibinfo {pages} {085009} (\bibinfo {year}
  {2011})}\BibitemShut {NoStop}%
\bibitem [{\citenamefont {Tey}\ \emph {et~al.}(2008)\citenamefont {Tey},
  \citenamefont {Chen}, \citenamefont {Aljunid}, \citenamefont {Chng},
  \citenamefont {Huber}, \citenamefont {Maslennikov},\ and\ \citenamefont
  {Kurtsiefer}}]{Tey2008}%
  \BibitemOpen
  \bibfield  {author} {\bibinfo {author} {\bibfnamefont {M.~K.}\ \bibnamefont
  {Tey}}, \bibinfo {author} {\bibfnamefont {Z.}~\bibnamefont {Chen}}, \bibinfo
  {author} {\bibfnamefont {S.~A.}\ \bibnamefont {Aljunid}}, \bibinfo {author}
  {\bibfnamefont {B.}~\bibnamefont {Chng}}, \bibinfo {author} {\bibfnamefont
  {F.}~\bibnamefont {Huber}}, \bibinfo {author} {\bibfnamefont
  {G.}~\bibnamefont {Maslennikov}}, \ and\ \bibinfo {author} {\bibfnamefont
  {C.}~\bibnamefont {Kurtsiefer}},\ }\href
  {http://dx.doi.org/10.1038/nphys1096} {\bibfield  {journal} {\bibinfo
  {journal} {Nat Phys}\ }\textbf {\bibinfo {volume} {4}},\ \bibinfo {pages}
  {924} (\bibinfo {year} {2008})}\BibitemShut {NoStop}%
\bibitem [{\citenamefont {Lee}\ \emph {et~al.}(2011)\citenamefont {Lee},
  \citenamefont {Chen}, \citenamefont {Eghlidi}, \citenamefont {Kukura},
  \citenamefont {Lettow}, \citenamefont {Renn}, \citenamefont {Sandoghdar},\
  and\ \citenamefont {G\"otzinger}}]{Lee2011}%
  \BibitemOpen
  \bibfield  {author} {\bibinfo {author} {\bibfnamefont {K.~G.}\ \bibnamefont
  {Lee}}, \bibinfo {author} {\bibfnamefont {X.~W.}\ \bibnamefont {Chen}},
  \bibinfo {author} {\bibfnamefont {H.}~\bibnamefont {Eghlidi}}, \bibinfo
  {author} {\bibfnamefont {P.}~\bibnamefont {Kukura}}, \bibinfo {author}
  {\bibfnamefont {R.}~\bibnamefont {Lettow}}, \bibinfo {author} {\bibfnamefont
  {A.}~\bibnamefont {Renn}}, \bibinfo {author} {\bibfnamefont {V.}~\bibnamefont
  {Sandoghdar}}, \ and\ \bibinfo {author} {\bibfnamefont {S.}~\bibnamefont
  {G\"otzinger}},\ }\href {http://dx.doi.org/10.1038/nphoton.2010.312}
  {\bibfield  {journal} {\bibinfo  {journal} {Nat Photon}\ }\textbf {\bibinfo
  {volume} {5}},\ \bibinfo {pages} {166} (\bibinfo {year} {2011})}\BibitemShut
  {NoStop}%
\bibitem [{\citenamefont {Maiwald}\ \emph {et~al.}(2012)\citenamefont
  {Maiwald}, \citenamefont {Golla}, \citenamefont {Fischer}, \citenamefont
  {Bader}, \citenamefont {Heugel}, \citenamefont {Chalopin}, \citenamefont
  {Sondermann},\ and\ \citenamefont {Leuchs}}]{Maiwald2012}%
  \BibitemOpen
  \bibfield  {author} {\bibinfo {author} {\bibfnamefont {R.}~\bibnamefont
  {Maiwald}}, \bibinfo {author} {\bibfnamefont {A.}~\bibnamefont {Golla}},
  \bibinfo {author} {\bibfnamefont {M.}~\bibnamefont {Fischer}}, \bibinfo
  {author} {\bibfnamefont {M.}~\bibnamefont {Bader}}, \bibinfo {author}
  {\bibfnamefont {S.}~\bibnamefont {Heugel}}, \bibinfo {author} {\bibfnamefont
  {B.}~\bibnamefont {Chalopin}}, \bibinfo {author} {\bibfnamefont
  {M.}~\bibnamefont {Sondermann}}, \ and\ \bibinfo {author} {\bibfnamefont
  {G.}~\bibnamefont {Leuchs}},\ }\href {\doibase 10.1103/PhysRevA.86.043431}
  {\bibfield  {journal} {\bibinfo  {journal} {Phys. Rev. A}\ }\textbf {\bibinfo
  {volume} {86}},\ \bibinfo {pages} {043431} (\bibinfo {year}
  {2012})}\BibitemShut {NoStop}%
\bibitem [{\citenamefont {Matsko}\ and\ \citenamefont
  {Ilchenko}(2006)}]{Matsko:2006aa}%
  \BibitemOpen
  \bibfield  {author} {\bibinfo {author} {\bibfnamefont {A.}~\bibnamefont
  {Matsko}}\ and\ \bibinfo {author} {\bibfnamefont {V.}~\bibnamefont
  {Ilchenko}},\ }\href {\doibase 10.1109/JSTQE.2005.862952} {\bibfield
  {journal} {\bibinfo  {journal} {Selected Topics in Quantum Electronics, IEEE
  Journal of}\ }\textbf {\bibinfo {volume} {12}},\ \bibinfo {pages} {3 }
  (\bibinfo {year} {2006})}\BibitemShut {NoStop}%
\bibitem [{\citenamefont {Spillane}\ \emph {et~al.}(2003)\citenamefont
  {Spillane}, \citenamefont {Kippenberg}, \citenamefont {Painter},\ and\
  \citenamefont {Vahala}}]{Spillane:2003aa}%
  \BibitemOpen
  \bibfield  {author} {\bibinfo {author} {\bibfnamefont {S.~M.}\ \bibnamefont
  {Spillane}}, \bibinfo {author} {\bibfnamefont {T.~J.}\ \bibnamefont
  {Kippenberg}}, \bibinfo {author} {\bibfnamefont {O.~J.}\ \bibnamefont
  {Painter}}, \ and\ \bibinfo {author} {\bibfnamefont {K.~J.}\ \bibnamefont
  {Vahala}},\ }\href@noop {} {\bibfield  {journal} {\bibinfo  {journal} {Phys.
  Rev. Lett.}\ }\textbf {\bibinfo {volume} {91}},\ \bibinfo {pages} {043902}
  (\bibinfo {year} {2003})}\BibitemShut {NoStop}%
\bibitem [{\citenamefont {Aoki}\ \emph {et~al.}(2006)\citenamefont {Aoki},
  \citenamefont {Dayan}, \citenamefont {Wilcut}, \citenamefont {Bowen},
  \citenamefont {Parkins}, \citenamefont {Kippenberg}, \citenamefont {Vahala},\
  and\ \citenamefont {Kimble}}]{Aoki:2006aa}%
  \BibitemOpen
  \bibfield  {author} {\bibinfo {author} {\bibfnamefont {T.}~\bibnamefont
  {Aoki}}, \bibinfo {author} {\bibfnamefont {B.}~\bibnamefont {Dayan}},
  \bibinfo {author} {\bibfnamefont {E.}~\bibnamefont {Wilcut}}, \bibinfo
  {author} {\bibfnamefont {W.~P.}\ \bibnamefont {Bowen}}, \bibinfo {author}
  {\bibfnamefont {A.~S.}\ \bibnamefont {Parkins}}, \bibinfo {author}
  {\bibfnamefont {T.~J.}\ \bibnamefont {Kippenberg}}, \bibinfo {author}
  {\bibfnamefont {K.~J.}\ \bibnamefont {Vahala}}, \ and\ \bibinfo {author}
  {\bibfnamefont {H.~J.}\ \bibnamefont {Kimble}},\ }\href {\doibase
  10.1038/nature05147} {\bibfield  {journal} {\bibinfo  {journal} {Nature}\
  }\textbf {\bibinfo {volume} {443}},\ \bibinfo {pages} {671} (\bibinfo {year}
  {2006})}\BibitemShut {NoStop}%
\bibitem [{\citenamefont {Park}\ \emph {et~al.}(2006)\citenamefont {Park},
  \citenamefont {Cook},\ and\ \citenamefont {Wang}}]{Park:2006aa}%
  \BibitemOpen
  \bibfield  {author} {\bibinfo {author} {\bibfnamefont {Y.}~\bibnamefont
  {Park}}, \bibinfo {author} {\bibfnamefont {A.}~\bibnamefont {Cook}}, \ and\
  \bibinfo {author} {\bibfnamefont {H.}~\bibnamefont {Wang}},\ }\href@noop {}
  {\bibfield  {journal} {\bibinfo  {journal} {Nano Lett.}\ }\textbf {\bibinfo
  {volume} {6}},\ \bibinfo {pages} {2075} (\bibinfo {year} {2006})}\BibitemShut
  {NoStop}%
\bibitem [{\citenamefont {Srinivasan}\ and\ \citenamefont
  {Painter}(2007{\natexlab{a}})}]{Srinivasan:2007aa}%
  \BibitemOpen
  \bibfield  {author} {\bibinfo {author} {\bibfnamefont {K.}~\bibnamefont
  {Srinivasan}}\ and\ \bibinfo {author} {\bibfnamefont {O.}~\bibnamefont
  {Painter}},\ }\href@noop {} {\bibfield  {journal} {\bibinfo  {journal}
  {Nature}\ }\textbf {\bibinfo {volume} {450}},\ \bibinfo {pages} {862}
  (\bibinfo {year} {2007}{\natexlab{a}})}\BibitemShut {NoStop}%
\bibitem [{\citenamefont {Dayan}\ \emph {et~al.}(2008)\citenamefont {Dayan},
  \citenamefont {Parkins}, \citenamefont {Aoki}, \citenamefont {Ostby},
  \citenamefont {Vahala},\ and\ \citenamefont {Kimble}}]{Dayan:2008aa}%
  \BibitemOpen
  \bibfield  {author} {\bibinfo {author} {\bibfnamefont {B.}~\bibnamefont
  {Dayan}}, \bibinfo {author} {\bibfnamefont {A.}~\bibnamefont {Parkins}},
  \bibinfo {author} {\bibfnamefont {T.}~\bibnamefont {Aoki}}, \bibinfo {author}
  {\bibfnamefont {E.}~\bibnamefont {Ostby}}, \bibinfo {author} {\bibfnamefont
  {K.}~\bibnamefont {Vahala}}, \ and\ \bibinfo {author} {\bibfnamefont
  {H.}~\bibnamefont {Kimble}},\ }\href@noop {} {\bibfield  {journal} {\bibinfo
  {journal} {Science}\ }\textbf {\bibinfo {volume} {319}},\ \bibinfo {pages}
  {1062} (\bibinfo {year} {2008})}\BibitemShut {NoStop}%
\bibitem [{\citenamefont {Aoki}\ \emph {et~al.}(2009)\citenamefont {Aoki},
  \citenamefont {Parkins}, \citenamefont {Alton}, \citenamefont {Regal},
  \citenamefont {Dayan}, \citenamefont {Ostby}, \citenamefont {Vahala},\ and\
  \citenamefont {Kimble}}]{Aoki:2009aa}%
  \BibitemOpen
  \bibfield  {author} {\bibinfo {author} {\bibfnamefont {T.}~\bibnamefont
  {Aoki}}, \bibinfo {author} {\bibfnamefont {A.~S.}\ \bibnamefont {Parkins}},
  \bibinfo {author} {\bibfnamefont {D.~J.}\ \bibnamefont {Alton}}, \bibinfo
  {author} {\bibfnamefont {C.~A.}\ \bibnamefont {Regal}}, \bibinfo {author}
  {\bibfnamefont {B.}~\bibnamefont {Dayan}}, \bibinfo {author} {\bibfnamefont
  {E.}~\bibnamefont {Ostby}}, \bibinfo {author} {\bibfnamefont {K.~J.}\
  \bibnamefont {Vahala}}, \ and\ \bibinfo {author} {\bibfnamefont {H.~J.}\
  \bibnamefont {Kimble}},\ }\href@noop {} {\bibfield  {journal} {\bibinfo
  {journal} {Phys. Rev. Lett.}\ }\textbf {\bibinfo {volume} {102}},\ \bibinfo
  {pages} {083601} (\bibinfo {year} {2009})}\BibitemShut {NoStop}%
\bibitem [{\citenamefont {Alton}\ \emph {et~al.}(2010)\citenamefont {Alton},
  \citenamefont {Stern}, \citenamefont {Aoki}, \citenamefont {Lee},
  \citenamefont {Ostby}, \citenamefont {Vahala},\ and\ \citenamefont
  {Kimble}}]{Alton:2010aa}%
  \BibitemOpen
  \bibfield  {author} {\bibinfo {author} {\bibfnamefont {D.~J.}\ \bibnamefont
  {Alton}}, \bibinfo {author} {\bibfnamefont {N.~P.}\ \bibnamefont {Stern}},
  \bibinfo {author} {\bibfnamefont {T.}~\bibnamefont {Aoki}}, \bibinfo {author}
  {\bibfnamefont {H.}~\bibnamefont {Lee}}, \bibinfo {author} {\bibfnamefont
  {E.}~\bibnamefont {Ostby}}, \bibinfo {author} {\bibfnamefont {K.~J.}\
  \bibnamefont {Vahala}}, \ and\ \bibinfo {author} {\bibfnamefont {H.~J.}\
  \bibnamefont {Kimble}},\ }\href@noop {} {\bibfield  {journal} {\bibinfo
  {journal} {Nat. Phys.}\ }\textbf {\bibinfo {volume} {7}},\ \bibinfo {pages}
  {159} (\bibinfo {year} {2010})}\BibitemShut {NoStop}%
\bibitem [{\citenamefont {Zhu}\ \emph {et~al.}(2010)\citenamefont {Zhu},
  \citenamefont {Ozdemir}, \citenamefont {Xiao}, \citenamefont {Li},
  \citenamefont {He}, \citenamefont {Chen},\ and\ \citenamefont
  {Yang}}]{Zhu:2010ab}%
  \BibitemOpen
  \bibfield  {author} {\bibinfo {author} {\bibfnamefont {J.}~\bibnamefont
  {Zhu}}, \bibinfo {author} {\bibfnamefont {S.~K.}\ \bibnamefont {Ozdemir}},
  \bibinfo {author} {\bibfnamefont {Y.-F.}\ \bibnamefont {Xiao}}, \bibinfo
  {author} {\bibfnamefont {L.}~\bibnamefont {Li}}, \bibinfo {author}
  {\bibfnamefont {L.}~\bibnamefont {He}}, \bibinfo {author} {\bibfnamefont
  {D.-R.}\ \bibnamefont {Chen}}, \ and\ \bibinfo {author} {\bibfnamefont
  {L.}~\bibnamefont {Yang}},\ }\href@noop {} {\bibfield  {journal} {\bibinfo
  {journal} {Nat. Photon.}\ }\textbf {\bibinfo {volume} {4}},\ \bibinfo {pages}
  {46} (\bibinfo {year} {2010})}\BibitemShut {NoStop}%
\bibitem [{\citenamefont {Vollmer}\ \emph {et~al.}(2008)\citenamefont
  {Vollmer}, \citenamefont {Arnold},\ and\ \citenamefont
  {Keng}}]{Vollmer:2008ab}%
  \BibitemOpen
  \bibfield  {author} {\bibinfo {author} {\bibfnamefont {F.}~\bibnamefont
  {Vollmer}}, \bibinfo {author} {\bibfnamefont {S.}~\bibnamefont {Arnold}}, \
  and\ \bibinfo {author} {\bibfnamefont {D.}~\bibnamefont {Keng}},\ }\href
  {\doibase 10.1073/pnas.0808988106} {\bibfield  {journal} {\bibinfo  {journal}
  {Proc. Natl. Acad. Sci. USA}\ }\textbf {\bibinfo {volume} {105}},\ \bibinfo
  {pages} {20701} (\bibinfo {year} {2008})}\BibitemShut {NoStop}%
\bibitem [{\citenamefont {Del'Haye}\ \emph {et~al.}(2007)\citenamefont
  {Del'Haye}, \citenamefont {Schliesser}, \citenamefont {Arcizet},
  \citenamefont {Wilken}, \citenamefont {Holzwarth},\ and\ \citenamefont
  {Kippenberg}}]{DelHaye:2007aa}%
  \BibitemOpen
  \bibfield  {author} {\bibinfo {author} {\bibfnamefont {P.}~\bibnamefont
  {Del'Haye}}, \bibinfo {author} {\bibfnamefont {A.}~\bibnamefont
  {Schliesser}}, \bibinfo {author} {\bibfnamefont {O.}~\bibnamefont {Arcizet}},
  \bibinfo {author} {\bibfnamefont {T.}~\bibnamefont {Wilken}}, \bibinfo
  {author} {\bibfnamefont {R.}~\bibnamefont {Holzwarth}}, \ and\ \bibinfo
  {author} {\bibfnamefont {T.~J.}\ \bibnamefont {Kippenberg}},\ }\href@noop {}
  {\bibfield  {journal} {\bibinfo  {journal} {Nature}\ }\textbf {\bibinfo
  {volume} {450}},\ \bibinfo {pages} {1214} (\bibinfo {year}
  {2007})}\BibitemShut {NoStop}%
\bibitem [{\citenamefont {F{\"u}rst}\ \emph {et~al.}(2011)\citenamefont
  {F{\"u}rst}, \citenamefont {Strekalov}, \citenamefont {Elser}, \citenamefont
  {Aiello}, \citenamefont {Andersen}, \citenamefont {Marquardt},\ and\
  \citenamefont {Leuchs}}]{Furst:2011aa}%
  \BibitemOpen
  \bibfield  {author} {\bibinfo {author} {\bibfnamefont {J.~U.}\ \bibnamefont
  {F{\"u}rst}}, \bibinfo {author} {\bibfnamefont {D.~V.}\ \bibnamefont
  {Strekalov}}, \bibinfo {author} {\bibfnamefont {D.}~\bibnamefont {Elser}},
  \bibinfo {author} {\bibfnamefont {A.}~\bibnamefont {Aiello}}, \bibinfo
  {author} {\bibfnamefont {U.~L.}\ \bibnamefont {Andersen}}, \bibinfo {author}
  {\bibfnamefont {C.}~\bibnamefont {Marquardt}}, \ and\ \bibinfo {author}
  {\bibfnamefont {G.}~\bibnamefont {Leuchs}},\ }\href@noop {} {\bibfield
  {journal} {\bibinfo  {journal} {Phys. Rev. Lett.}\ }\textbf {\bibinfo
  {volume} {106}},\ \bibinfo {pages} {113901} (\bibinfo {year}
  {2011})}\BibitemShut {NoStop}%
\bibitem [{\citenamefont {F{\"o}rtsch}\ \emph {et~al.}(2012)\citenamefont
  {F{\"o}rtsch}, \citenamefont {F{\"u}rst}, \citenamefont {Wittmann},
  \citenamefont {Strekalov}, \citenamefont {Aiello}, \citenamefont {Chekhova},
  \citenamefont {Silberhorn}, \citenamefont {Leuchs},\ and\ \citenamefont
  {Marquardt}}]{Fortsch:2012aa}%
  \BibitemOpen
  \bibfield  {author} {\bibinfo {author} {\bibfnamefont {M.}~\bibnamefont
  {F{\"o}rtsch}}, \bibinfo {author} {\bibfnamefont {J.}~\bibnamefont
  {F{\"u}rst}}, \bibinfo {author} {\bibfnamefont {C.}~\bibnamefont {Wittmann}},
  \bibinfo {author} {\bibfnamefont {D.}~\bibnamefont {Strekalov}}, \bibinfo
  {author} {\bibfnamefont {A.}~\bibnamefont {Aiello}}, \bibinfo {author}
  {\bibfnamefont {M.~V.}\ \bibnamefont {Chekhova}}, \bibinfo {author}
  {\bibfnamefont {C.}~\bibnamefont {Silberhorn}}, \bibinfo {author}
  {\bibfnamefont {G.}~\bibnamefont {Leuchs}}, \ and\ \bibinfo {author}
  {\bibfnamefont {C.}~\bibnamefont {Marquardt}},\ }\href@noop {} {\bibfield
  {journal} {\bibinfo  {journal} {arXiv:1204.3056}\ } (\bibinfo {year}
  {2012})}\BibitemShut {NoStop}%
\bibitem [{\citenamefont {Kippenberg}\ and\ \citenamefont
  {Vahala}(2008)}]{Kippenberg:2008aa}%
  \BibitemOpen
  \bibfield  {author} {\bibinfo {author} {\bibfnamefont {T.~J.}\ \bibnamefont
  {Kippenberg}}\ and\ \bibinfo {author} {\bibfnamefont {K.~J.}\ \bibnamefont
  {Vahala}},\ }\href {\doibase 10.1126/science.1156032} {\bibfield  {journal}
  {\bibinfo  {journal} {Science}\ }\textbf {\bibinfo {volume} {321}},\ \bibinfo
  {pages} {1172} (\bibinfo {year} {2008})}\BibitemShut {NoStop}%
\bibitem [{\citenamefont {Bahl}\ \emph {et~al.}(2012)\citenamefont {Bahl},
  \citenamefont {Tomes}, \citenamefont {Marquardt},\ and\ \citenamefont
  {Carmon}}]{Bahl:2012aa}%
  \BibitemOpen
  \bibfield  {author} {\bibinfo {author} {\bibfnamefont {G.}~\bibnamefont
  {Bahl}}, \bibinfo {author} {\bibfnamefont {M.}~\bibnamefont {Tomes}},
  \bibinfo {author} {\bibfnamefont {F.}~\bibnamefont {Marquardt}}, \ and\
  \bibinfo {author} {\bibfnamefont {T.}~\bibnamefont {Carmon}},\ }\href@noop {}
  {\bibfield  {journal} {\bibinfo  {journal} {Nat. Phys.}\ }\textbf {\bibinfo
  {volume} {8}},\ \bibinfo {pages} {203} (\bibinfo {year} {2012})}\BibitemShut
  {NoStop}%
\bibitem [{\citenamefont {He}\ \emph {et~al.}(2011)\citenamefont {He},
  \citenamefont {Ozdemir}, \citenamefont {Zhu}, \citenamefont {Kim},\ and\
  \citenamefont {Yang}}]{He:2011aa}%
  \BibitemOpen
  \bibfield  {author} {\bibinfo {author} {\bibfnamefont {L.}~\bibnamefont
  {He}}, \bibinfo {author} {\bibfnamefont {S.~K.}\ \bibnamefont {Ozdemir}},
  \bibinfo {author} {\bibfnamefont {J.}~\bibnamefont {Zhu}}, \bibinfo {author}
  {\bibfnamefont {W.}~\bibnamefont {Kim}}, \ and\ \bibinfo {author}
  {\bibfnamefont {L.}~\bibnamefont {Yang}},\ }\href@noop {} {\bibfield
  {journal} {\bibinfo  {journal} {Nat. Nano.}\ }\textbf {\bibinfo {volume}
  {6}},\ \bibinfo {pages} {428} (\bibinfo {year} {2011})}\BibitemShut {NoStop}%
\bibitem [{\citenamefont {Domokos}\ \emph {et~al.}(2000)\citenamefont
  {Domokos}, \citenamefont {Gangl},\ and\ \citenamefont
  {Ritsch}}]{Domokos:2000aa}%
  \BibitemOpen
  \bibfield  {author} {\bibinfo {author} {\bibfnamefont {P.}~\bibnamefont
  {Domokos}}, \bibinfo {author} {\bibfnamefont {M.}~\bibnamefont {Gangl}}, \
  and\ \bibinfo {author} {\bibfnamefont {H.}~\bibnamefont {Ritsch}},\
  }\href@noop {} {\bibfield  {journal} {\bibinfo  {journal} {Opt. Commun.}\
  }\textbf {\bibinfo {volume} {185}},\ \bibinfo {pages} {115} (\bibinfo {year}
  {2000})}\BibitemShut {NoStop}%
\bibitem [{\citenamefont {Rosenblit}\ \emph {et~al.}(2004)\citenamefont
  {Rosenblit}, \citenamefont {Horak}, \citenamefont {Helsby},\ and\
  \citenamefont {Folman}}]{Rosenblit:2004aa}%
  \BibitemOpen
  \bibfield  {author} {\bibinfo {author} {\bibfnamefont {M.}~\bibnamefont
  {Rosenblit}}, \bibinfo {author} {\bibfnamefont {P.}~\bibnamefont {Horak}},
  \bibinfo {author} {\bibfnamefont {S.}~\bibnamefont {Helsby}}, \ and\ \bibinfo
  {author} {\bibfnamefont {R.}~\bibnamefont {Folman}},\ }\href@noop {}
  {\bibfield  {journal} {\bibinfo  {journal} {Phys. Rev. A}\ }\textbf {\bibinfo
  {volume} {70}},\ \bibinfo {pages} {053808} (\bibinfo {year}
  {2004})}\BibitemShut {NoStop}%
\bibitem [{\citenamefont {Srinivasan}\ and\ \citenamefont
  {Painter}(2007{\natexlab{b}})}]{Srinivasan:2007ab}%
  \BibitemOpen
  \bibfield  {author} {\bibinfo {author} {\bibfnamefont {K.}~\bibnamefont
  {Srinivasan}}\ and\ \bibinfo {author} {\bibfnamefont {O.}~\bibnamefont
  {Painter}},\ }\href {\doibase 10.1103/PhysRevA.75.023814} {\bibfield
  {journal} {\bibinfo  {journal} {Phys. Rev. A}\ }\textbf {\bibinfo {volume}
  {75}},\ \bibinfo {pages} {023814} (\bibinfo {year}
  {2007}{\natexlab{b}})}\BibitemShut {NoStop}%
\bibitem [{\citenamefont {Shen}\ and\ \citenamefont {Fan}(2009)}]{Shen:2009ab}%
  \BibitemOpen
  \bibfield  {author} {\bibinfo {author} {\bibfnamefont {J.~T.}\ \bibnamefont
  {Shen}}\ and\ \bibinfo {author} {\bibfnamefont {S.}~\bibnamefont {Fan}},\
  }\href@noop {} {\bibfield  {journal} {\bibinfo  {journal} {Phys. Rev. A}\
  }\textbf {\bibinfo {volume} {79}},\ \bibinfo {pages} {023838} (\bibinfo
  {year} {2009})}\BibitemShut {NoStop}%
\bibitem [{\citenamefont {Stern}\ \emph {et~al.}(2011)\citenamefont {Stern},
  \citenamefont {Alton},\ and\ \citenamefont {Kimble}}]{Stern:2011aa}%
  \BibitemOpen
  \bibfield  {author} {\bibinfo {author} {\bibfnamefont {N.~P.}\ \bibnamefont
  {Stern}}, \bibinfo {author} {\bibfnamefont {D.~J.}\ \bibnamefont {Alton}}, \
  and\ \bibinfo {author} {\bibfnamefont {H.~J.}\ \bibnamefont {Kimble}},\
  }\href@noop {} {\bibfield  {journal} {\bibinfo  {journal} {New J. Phys.}\
  }\textbf {\bibinfo {volume} {13}},\ \bibinfo {pages} {085004} (\bibinfo
  {year} {2011})}\BibitemShut {NoStop}%
\bibitem [{\citenamefont {Sumetsky}(2004)}]{Sumetsky2004}%
  \BibitemOpen
  \bibfield  {author} {\bibinfo {author} {\bibfnamefont {M.}~\bibnamefont
  {Sumetsky}},\ }\href@noop {} {\bibfield  {journal} {\bibinfo  {journal} {Opt.
  Lett.}\ }\textbf {\bibinfo {volume} {29}},\ \bibinfo {pages} {8} (\bibinfo
  {year} {2004})}\BibitemShut {NoStop}%
\bibitem [{\citenamefont {Louyer}\ \emph {et~al.}(2005)\citenamefont {Louyer},
  \citenamefont {Meschede},\ and\ \citenamefont
  {Rauschenbeutel}}]{Louyer:2005aa}%
  \BibitemOpen
  \bibfield  {author} {\bibinfo {author} {\bibfnamefont {Y.}~\bibnamefont
  {Louyer}}, \bibinfo {author} {\bibfnamefont {D.}~\bibnamefont {Meschede}}, \
  and\ \bibinfo {author} {\bibfnamefont {A.}~\bibnamefont {Rauschenbeutel}},\
  }\href@noop {} {\bibfield  {journal} {\bibinfo  {journal} {Phys. Rev. A}\
  }\textbf {\bibinfo {volume} {72}},\ \bibinfo {pages} {031801} (\bibinfo
  {year} {2005})}\BibitemShut {NoStop}%
\bibitem [{\citenamefont {P{\"o}llinger}\ \emph {et~al.}(2009)\citenamefont
  {P{\"o}llinger}, \citenamefont {O'Shea}, \citenamefont {Warken},\ and\
  \citenamefont {Rauschenbeutel}}]{Pollinger:2009aa}%
  \BibitemOpen
  \bibfield  {author} {\bibinfo {author} {\bibfnamefont {M.}~\bibnamefont
  {P{\"o}llinger}}, \bibinfo {author} {\bibfnamefont {D.}~\bibnamefont
  {O'Shea}}, \bibinfo {author} {\bibfnamefont {F.}~\bibnamefont {Warken}}, \
  and\ \bibinfo {author} {\bibfnamefont {A.}~\bibnamefont {Rauschenbeutel}},\
  }\href@noop {} {\bibfield  {journal} {\bibinfo  {journal} {Phys. Rev. Lett.}\
  }\textbf {\bibinfo {volume} {103}},\ \bibinfo {pages} {053901} (\bibinfo
  {year} {2009})}\BibitemShut {NoStop}%
\bibitem [{\citenamefont {Braginsky}\ \emph {et~al.}(1989)\citenamefont
  {Braginsky}, \citenamefont {Gorodetsky},\ and\ \citenamefont
  {Ilchenko}}]{Braginsky:1989aa}%
  \BibitemOpen
  \bibfield  {author} {\bibinfo {author} {\bibfnamefont {V.}~\bibnamefont
  {Braginsky}}, \bibinfo {author} {\bibfnamefont {M.}~\bibnamefont
  {Gorodetsky}}, \ and\ \bibinfo {author} {\bibfnamefont {V.}~\bibnamefont
  {Ilchenko}},\ }\href {\doibase 10.1016/0375-9601(89)90912-2} {\bibfield
  {journal} {\bibinfo  {journal} {Physics Letters A}\ }\textbf {\bibinfo
  {volume} {137}},\ \bibinfo {pages} {393 } (\bibinfo {year}
  {1989})}\BibitemShut {NoStop}%
\bibitem [{\citenamefont {Armani}\ \emph {et~al.}(2003)\citenamefont {Armani},
  \citenamefont {Kippenberg}, \citenamefont {Spillane},\ and\ \citenamefont
  {Vahala}}]{Armani:2003aa}%
  \BibitemOpen
  \bibfield  {author} {\bibinfo {author} {\bibfnamefont {D.~K.}\ \bibnamefont
  {Armani}}, \bibinfo {author} {\bibfnamefont {T.~J.}\ \bibnamefont
  {Kippenberg}}, \bibinfo {author} {\bibfnamefont {S.~M.}\ \bibnamefont
  {Spillane}}, \ and\ \bibinfo {author} {\bibfnamefont {K.~J.}\ \bibnamefont
  {Vahala}},\ }\href@noop {} {\bibfield  {journal} {\bibinfo  {journal}
  {Nature}\ }\textbf {\bibinfo {volume} {421}},\ \bibinfo {pages} {925}
  (\bibinfo {year} {2003})}\BibitemShut {NoStop}%
\bibitem [{\citenamefont {P{\"o}llinger}\ and\ \citenamefont
  {Rauschenbeutel}(2010)}]{Pollinger:2010aa}%
  \BibitemOpen
  \bibfield  {author} {\bibinfo {author} {\bibfnamefont {M.}~\bibnamefont
  {P{\"o}llinger}}\ and\ \bibinfo {author} {\bibfnamefont {A.}~\bibnamefont
  {Rauschenbeutel}},\ }\href@noop {} {\bibfield  {journal} {\bibinfo  {journal}
  {Opt. Express}\ }\textbf {\bibinfo {volume} {18}},\ \bibinfo {pages} {17764}
  (\bibinfo {year} {2010})}\BibitemShut {NoStop}%
\bibitem [{\citenamefont {Axelrod}\ \emph {et~al.}(1984)\citenamefont
  {Axelrod}, \citenamefont {Burghardt},\ and\ \citenamefont
  {Thompson}}]{Axelrod:1984aa}%
  \BibitemOpen
  \bibfield  {author} {\bibinfo {author} {\bibfnamefont {D.}~\bibnamefont
  {Axelrod}}, \bibinfo {author} {\bibfnamefont {T.}~\bibnamefont {Burghardt}},
  \ and\ \bibinfo {author} {\bibfnamefont {N.}~\bibnamefont {Thompson}},\
  }\href@noop {} {\bibfield  {journal} {\bibinfo  {journal} {Annu. Rev.
  Biophys. Bio.}\ }\textbf {\bibinfo {volume} {13}},\ \bibinfo {pages} {247}
  (\bibinfo {year} {1984})}\BibitemShut {NoStop}%
\bibitem [{\citenamefont {Kawalec}\ \emph {et~al.}(2007)\citenamefont
  {Kawalec}, \citenamefont {J\'{o}zefowski}, \citenamefont {Fiutowski},
  \citenamefont {Kasprowicz},\ and\ \citenamefont {Dohnalik}}]{Kawalec:2007aa}%
  \BibitemOpen
  \bibfield  {author} {\bibinfo {author} {\bibfnamefont {T.}~\bibnamefont
  {Kawalec}}, \bibinfo {author} {\bibfnamefont {L.}~\bibnamefont
  {J\'{o}zefowski}}, \bibinfo {author} {\bibfnamefont {J.}~\bibnamefont
  {Fiutowski}}, \bibinfo {author} {\bibfnamefont {M.}~\bibnamefont
  {Kasprowicz}}, \ and\ \bibinfo {author} {\bibfnamefont {T.}~\bibnamefont
  {Dohnalik}},\ }\href {\doibase 10.1016/j.optcom.2007.02.042} {\bibfield
  {journal} {\bibinfo  {journal} {Optics Communications}\ }\textbf {\bibinfo
  {volume} {274}},\ \bibinfo {pages} {341 } (\bibinfo {year}
  {2007})}\BibitemShut {NoStop}%
\bibitem [{\citenamefont {O'Shea}\ \emph {et~al.}(2010)\citenamefont {O'Shea},
  \citenamefont {Rettenmaier},\ and\ \citenamefont
  {Rauschenbeutel}}]{OShea:2010aa}%
  \BibitemOpen
  \bibfield  {author} {\bibinfo {author} {\bibfnamefont {D.}~\bibnamefont
  {O'Shea}}, \bibinfo {author} {\bibfnamefont {A.}~\bibnamefont {Rettenmaier}},
  \ and\ \bibinfo {author} {\bibfnamefont {A.}~\bibnamefont {Rauschenbeutel}},\
  }\href@noop {} {\bibfield  {journal} {\bibinfo  {journal} {Appl. Phys. B}\
  }\textbf {\bibinfo {volume} {99}},\ \bibinfo {pages} {623} (\bibinfo {year}
  {2010})}\BibitemShut {NoStop}%
\bibitem [{\citenamefont {O'Shea}\ \emph {et~al.}(2011)\citenamefont {O'Shea},
  \citenamefont {Junge}, \citenamefont {P{\"o}llinger}, \citenamefont
  {Vogler},\ and\ \citenamefont {Rauschenbeutel}}]{OShea:2011ab}%
  \BibitemOpen
  \bibfield  {author} {\bibinfo {author} {\bibfnamefont {D.}~\bibnamefont
  {O'Shea}}, \bibinfo {author} {\bibfnamefont {C.}~\bibnamefont {Junge}},
  \bibinfo {author} {\bibfnamefont {M.}~\bibnamefont {P{\"o}llinger}}, \bibinfo
  {author} {\bibfnamefont {A.}~\bibnamefont {Vogler}}, \ and\ \bibinfo {author}
  {\bibfnamefont {A.}~\bibnamefont {Rauschenbeutel}},\ }\href@noop {}
  {\bibfield  {journal} {\bibinfo  {journal} {Appl. Phys. B}\ }\textbf
  {\bibinfo {volume} {105}},\ \bibinfo {pages} {129} (\bibinfo {year}
  {2011})}\BibitemShut {NoStop}%
\bibitem [{\citenamefont {Junge}\ \emph {et~al.}(2011)\citenamefont {Junge},
  \citenamefont {Nickel}, \citenamefont {O'Shea},\ and\ \citenamefont
  {Rauschenbeutel}}]{Junge:2011aa}%
  \BibitemOpen
  \bibfield  {author} {\bibinfo {author} {\bibfnamefont {C.}~\bibnamefont
  {Junge}}, \bibinfo {author} {\bibfnamefont {S.}~\bibnamefont {Nickel}},
  \bibinfo {author} {\bibfnamefont {D.}~\bibnamefont {O'Shea}}, \ and\ \bibinfo
  {author} {\bibfnamefont {A.}~\bibnamefont {Rauschenbeutel}},\ }\href@noop {}
  {\bibfield  {journal} {\bibinfo  {journal} {Opt. Lett.}\ }\textbf {\bibinfo
  {volume} {36}},\ \bibinfo {pages} {3488} (\bibinfo {year}
  {2011})}\BibitemShut {NoStop}%
\end{thebibliography}%

\section*{Supplementary Information}

\subsection*{Atom source}
In our experiment, we prepare a cloud of about $5\times10^7$ laser-cooled $^{85}$Rb atoms in a magneto-optical trap and launch it upwards by means of a moving molasses \cite{OShea:2011ab}. The bottle microresonator is approximately located at the turning point of the ballistic trajectory of the atomic cloud. The atom cloud overlaps with the bottle microresonator for about 50~ms. During this time interval, a small number of individual atoms will enter into the evanescent field of the WGM at random arrival times.  The temperature of the atom cloud is about 5~$\mu$K, which yields a thermal velocity on the order of a few cm/s. In conjunction with the decay length of the evanescent field of the bottle microresonator of about 118~nm, this gives rise to expected average interaction times of a few microseconds, in good agreement with the observed transits. We checked that the number of atom transits  depends strictly linearly on the density of the atomic cloud. This allows us to conclude that the observed coupling events are indeed transits of individual single atoms through the resonator field.

\subsection*{Real-time atom detection and spectroscopy}
The arrival time of atoms at the bottle microresonator is inherently random. Thus, an active detection scheme is applied, that can detect the presence of an atom in the cavity mode in real time. For this purpose, we set the coupling fibre to critical coupling and apply a detection light field with a flux of $1.2\times10^7$ photons/s, corresponding to a mean photon number of 0.18 in the resonator mode. We continuously monitor the coupling fibre transmission using single photon counting avalanche photodiodes. A trigger and control system, based on a fast field programmable gate array, registers the photon detection events. As soon as the number of detected photons within a predetermined time interval $\Delta t_1=1.2$~$\mu$s exceeds a user-defined threshold value $\eta_1=6$ (TM) or $\eta_1=4$ (TE), respectively, the system generates a trigger. This in turn operates a fast optical switch, that turns off the detection light and switches on a spectroscopy field that can be set to any relevant detuning  in advance. The spectroscopy light is applied for a predetermined time interval ranging from 100~ns to a few microseconds. We then switch back to the probe light in order to check if the atom is still present, using a detection time interval of $\Delta t_2=1$~$\mu$s and a threshold value of $\eta_2=2$ photons. Events where the atom was lost are discarded. For the spectra presented in this work, we typically average the transmission over about 1000 atom transits for each detuning of the spectroscopy light.

\subsection*{Bottle resonator modes}
In contrast to other WGM resonators, bottle-resonators sustain a large quantity of higher order axial modes \cite{Pollinger:2009aa}, which strongly differ by the intrinsic scattering rate between the two counter-propagating modes.  We estimate the axial mode number by imaging the scattered light from the empty resonator, which allows us to roughly infer the spatial extension of the cavity field. In order to achieve a high coupling strength in our experiment, we select high-$Q$ modes with minimal axial mode number, which feature a low intrinsic scattering rate at the same time. We infer the latter from the intrinsic splitting of the empty resonator transmission spectrum in the undercoupled regime, see Figs. 2 \textbf{c} and 2 \textbf{d}. For the modes investigated here, we estimate an axial mode number of $q\lesssim2$ ($q\lesssim4$) and an intrinsic mode splitting of smaller than $1$ MHz ($8$ MHz) for the TM (TE) mode.

\subsection*{Polarization of evanescent fields}
When light undergoes total internal reflection at a dielectric interface with refractive indices $n_1>n_2$, the amplitude of the two evanescent electric field components of TM waves is given by \cite{Axelrod:1984aa,Kawalec:2007aa}:
\begin{eqnarray}
E_r&=&E_0\frac{2\cos\theta\sin\theta}{(n^4\cos^2\theta+\sin^2\theta-n^2)^{1/2}}\\
E_\varphi&=&iE_0\frac{2\cos\theta(\sin^2\theta-n^2)^{1/2}}{(n^4\cos^2\theta+\sin^2\theta-n^2)^{1/2}},
\end{eqnarray}
where $\theta$ is angle of incidence and $n=n_2/n_1<1$ is the ratio of the two refractive indices. The optical modes of WGM we are interested in propagate close to the surface, which results in a large angle of incidence $\theta\approx\pi/2$. The ratio of the two electric field components at the boundary then is
\begin{equation}
\left|\frac{E_\varphi}{E_r}\right|=\sqrt{1-n^2},
\end{equation}
and is approximately constant in the evanescent field.

\subsection*{Modeling the transmission spectra}
For the description of CQED in our bottle microresonator system, we start with the Janyes-Cummings Hamiltonian describing the interaction of a two-level atom with two optical modes, as used in references \cite{Domokos:2000aa,Rosenblit:2004aa,Aoki:2006aa,Srinivasan:2007aa,Shen:2009ab}, for example. We extend this model to include the full Zeeman substructure of the atom and the polarization of the optical modes. Intrinsic scattering between the two counter-propagating bottle mode is experimentally found to be smaller than atom-resonator coupling rate $g$ as well as the decay rate of the resonator field and was thus neglected in the modeling the presented spectra. Since $g$ depends on the distance between the atom and the resonator surface in our experiment, it changes during the atom transits and from shot to shot. We account for this distribution of coupling strengths by fitting an averaged spectrum to the transmission data. For the averaging, we assume the coupling strengths to be normally distributed within an interval ranging from $g/2\pi=7.5$~MHz to $30$~MHz. The latter value corresponds to the expected maximum coupling strength at an atom--resonator distance of 50~nm. For smaller distances, the atom--surface interaction leads to significant shifts of the atomic transition frequencies which effectively inhibit the atom--resonator coupling.

\end{document}